  \documentstyle[12pt]{article}
  
\catcode`@=11
\def\secteqno{\@addtoreset{equation}{section}%
\def\theequation{\thesection.\arabic{equation}}}
\catcode`@=12
\secteqno
\newcommand{\be}{\begin{equation}}
\newcommand{\ee}{\end{equation}}
\newcommand{\bea}{\begin{eqnarray}}
\newcommand{\eea}{\end{eqnarray}}
\newcommand{\bref}[1]{(\ref{#1})}
\newcommand{\nn}{\nonumber}

\def\CL{{\cal L}}

\def\vsej{\vskip 4mm} 

\newcommand{\slSigma}{/ {\hskip-0.27cm{\Sigma}}}
\newcommand{\slP}{/ {\hskip-0.27cm{P}}}

\newcommand{\slX}{/ {\hskip-0.27cm{X}}}
\newcommand{\slp}{/ {\hskip-0.27cm{p}}}

\newcommand{\slbL}{/ {\hskip-0.27cm{\bf L}}}

\begin{document}
\vfill
\vbox{
\hfill July, 2000 \null\par
\hfill KEK-TH-702\null\par
\hfill TOHO-FP-0065
\par
\hfill YITP-00-34}\null
\vskip 20mm
\begin{center}
{\Large\bf Nondegenerate Super-Anti-de Sitter Algebra }\par
\vskip 6mm
{\Large\bf and a Superstring Action}
\vskip 20mm
{\large Machiko\ Hatsuda\footnote{mhatsuda@post.kek.jp},~Kiyoshi~Kamimura\footnote{kamimura@ph.sci.toho-u.ac.jp}$^\dagger$~and~~Makoto\ Sakaguchi\footnote{sakaguch@yukawa.kyoto-u.ac.jp
} $^{\dagger\dagger}$}\par
\medskip
Theory Division,\ High Energy Accelerator Research Organization (KEK),\\
\ Tsukuba,\ Ibaraki,\ 305-0801, Japan \\
$\dagger$ Department of Physics,\ Toho University,
\ Funabashi,\ 274-8510, Japan\\
$\dagger\dagger$ Yukawa Institute for Theoretical Physics,\ Kyoto University,\\
\ Sakyo-ku,\ Kyoto,\ 606-8502, Japan\\
\medskip
\vskip 10mm
\end{center}
\vskip 10mm
\begin{abstract}
We construct an Anti-de Sitter(AdS) algebra in a nondegenerate superspace.
Based on this algebra we construct a 
covariant kappa-symmetric superstring action,
and we examine its dynamics:
Although this action reduces to the usual Green-Schwarz superstring action in flat limit,
 the auxiliary fermionic coordinates
of the nondegenerate superspace becomes dynamical in the AdS background.
\end{abstract} 
\noindent{\it PACS:} 11.30.Pb;11.17.+y;11.25.-w \par\noindent
{\it Keywords:}  Superalgebra; SUSY central extension; BPS states;
 D-brane; Anti-de Sitter;
\par
\newpage
\setcounter{page}{1}
\parskip=7pt
%%%%%%%%%%%%%%%%%%%%%%%%%%%%%%%%%%%%%%%%%%%%%%%%%%%%%%%%%%%%%%%%%%%%%%%%%
\section{ Introduction}\par
\indent

After the conjecture of AdS/CFT correspondence \cite{Mal},
supersymmetric graded algebras based on \cite{HLS} have been reexamined 
\cite{Pro} and superstring actions in AdS spaces   
have been studied intensively \cite{MeTsy,Ram,Berk}.
In these references the superstring actions are constructed
as $\sigma$-models on coset superspaces of suitable graded
algebras.
These superalgebras are degenerate in a sense that 
nondegenerate metric for fermionic sector can not be defined 
using only one kind of supergenerator. 
Green showed that a nondegenerate superspace can be defined
by introducing a fermionic central charge \cite{Green} 
in a flat background.
It is an interesting  issue to examine how the fermionic ``central extension" 
is incorporated with the super-AdS algebra,
since the ``fermionic center" can not stay 
as a center anymore in the AdS space.
In this paper we discuss on the issue of 
``nondegenerate super-AdS algebra".

In the usual Green-Schwarz (GS) 
superstring \cite{GSSUST} the Wess-Zumino action
is pseudo (quasi) invariant under supersymmetry transformations.
The Noether charges of the supersymmetry acquire additional contributions
from the surface term. As a result the SUSY algebra contains anomaly or
topological terms \cite{azcTow} 
which play important roles in discussions of the 
BPS properties \cite{WitOl}.
On the other hand Siegel has shown that
the Wess-Zumino action of the superstring can be obtained 
by a simple bilinear combination of supercovariant Maurer-Cartan 1-forms
 for the nondegenerate 
supertranslation algebra.
This method overcomes the difficulty of the Wess-Zumino action on the
random lattice, then it gives a second quantized particle superfield theory
 \cite{Siegel}.
The Wess-Zumino action obtained in this way is an element of a trivial 
class of Chevalley-Eilenberg cohomology on the nondegenerate superspace
\cite{azcTow} which is manifestly SUSY invariant. 
In the AdS space the bilinear Wess-Zumino (WZ) 
action for the usual GS superstring 
could be written formally, but the bilinear WZ action
 contains the AdS radius parameter then the WZ term vanishes in the flat limit. 
 In order to get the bilinear WZ action, 
 the nondegenerate superalgebra which includes the new spinor charge
 must be constructed.

In this nondegenerate approach no topological term appears in the SUSY algebra.
Information of topological charge is contained in the anti-commutator
 of fermionic constraints 
\cite{WScm,azcTow2} both for the usual GS superstring and for the ``nondegenerate" 
superstring. In the anti-commutator of the fermionic constraints the topological term makes a half of 
fermionic constraints to be first class
which generate kappa-symmetry. 
In the nondegenerate superstring action there appear  
additional fermionic variables 
associated with new fermionic charges. 
In flat space the Lagrangian of the nondegenerate superstring 
coincides with the Lagrangian of the usual GS superstring
up to surface term \cite{Siegel}.
Since the new variables appear in the surface term
they lead to additional constraints which are almost trivially solved
and the constraint set is reduced to the usual one of
the GS superstring \cite{AHKT}.
There are, however, subtle differences in their SUSY algebra
and constraint algebra 
caused from different canonical variables. 
Arbitrary p-brane can be also described in
 the nondegenerate superalgebra approach 
 in the flat background  
\cite{Sez,Sakag,Sakag2,azc1}.

The equivalence between the usual GS superstring and 
``the nondegenerate superstring" is no longer hold in general background 
such as the AdS space,
because the fermionic charge is no more a center.
Criteria which were used for constructing the GS superstring actions in AdS spaces \cite{MeTsy,Ram}
are followings \cite{MeTsy}
:
\par
$\ast$  the standard $\sigma$-model bosonic part 
        with AdS structure as a target space\par
 $\ast$ the global AdS supersymmetry \par
 $\ast$ the local $\kappa$-symmetry\par
 $\ast$ reducing to the standard Green-Schwarz superstring action in the 
flat limit\par\noindent
In this paper we will construct a superstring action satisfying 
these criteria
based on a nondegenerate super-AdS algebra
\footnote{A superstring action in N=2 AdS$_2\times$S$^2$ was suggested as the nondegenerate superalgebra form in the reference \cite{Berk}. The existence of corresponding nondegenerate superalgebra itself is a nontrivial problem. 
We thank Nathan Berkovits for useful discussions
about these issues.}.
In order to make concrete calculation simpler we focus on
the AdS space in this paper rather than realistic AdS$\times$S target spaces.
We will examine the difference from the usual GS superstring,
such as relation of two supercharges,
anti-commutators of the fermionic constraints,
and dynamical modes.

This paper is organized as follows. In section 2, we introduce 
the nondegenerate super-AdS algebra and we give scale dimensions 
to generators in such a way that this algebra reduces into
the nondegenerate supertranslation algebra in the flat limit.
In section 3, we calculate Maurer-Cartan 1-forms of a coset $G/H$
where $G=$ (the nondegenerate super-AdS group) and $H=$ (Lorentz group).
In section 4, by using these expressions we construct 
a superstring action in the nondegenerate AdS space, and we examine 
differences between this nondegenerate superstring 
approach and the usual GS superstring approach both for 
the flat limit case and for the AdS case.  

%%%%%%%%%%%%%%%%%%%%%%%%%%%%%%%%%%%%%%%%%%%%%%%%%%%%%%%%%%%%%%%%%%%%%%%%
\section{ Nondegenerate Super-Anti-de Sitter Algebra}\par
\indent

We show that the following algebra is given as the nondegenerate 
super-anti de Sitter algebra in $d$ dimensions:
\bea
\begin{array}{lclcl}
\left[ J_{mn},J_{lk} \right]=\eta_{[k|[m}J_{n]|l]}
&,&
\left[  Q ,J_{mn}\right]=-\frac{1}{2}Q\Gamma_{mn}&,&  
\{ Q, Q\}=-2iC\slP%{\cal P}_+
\\&&&&\\
\left[ P_{m},J_{lk} \right]=\eta_{m[l}P_{k]}&,&
\left[ Z ,J_{mn}\right]=-\frac{1}{2}Z\Gamma_{mn}&,&  
\{Q,Z\}=C\Gamma^{mn}J_{mn}%{\cal P}_-
\\&&&&\\
\left[ P_{m},P_{l} \right]=J_{ml}&,&
\left[ Q ,P_{m}\right]=-\frac{i}{2}Z\Gamma_{m}&,&  
\{Z,Z\}=2iC\slP%{\cal P}_-
\\&&&&\\
&&
\left[ Z ,P_{m}\right]=\frac{i}{2}Q\Gamma_{m}
&&\end{array}\label{SAdS}
\eea 
where $J_{mn}$\footnote{$\eta_{[k|[m}J_{n]|l]}=\eta_{km}J_{nl}-\eta_{kn}J_{ml}
-\eta_{lm}J_{nk}+\eta_{ln}J_{mk}$ }
, $P_m$ are AdS$_d$ generators, and 
$Q_\alpha$ and $Z_\alpha$ are $d$-dimensional Majorana spinor
generators.
This algebra exists for
 $d=3$ %dimensional space ($_m=0,1,d-1$) 
 and its bosonic 
 symmetry group is
$SO(2,2)$ where the Majorana representation exists
with the anti-symmetric charge conjugation matrix, $C_{(-)}^T=-C_{(-)}$.

A flat limit is realized  by giving following scale dimensions to the generators as
\bea
P_m\to RP_m,~
Q_\alpha\to R^{1/2}Q_\alpha,~ Z_\alpha\to R^{3/2}Z_\alpha~ 
 {\rm and}~ J_{mn} \to J_{mn}\label{scaleAdS}
 \eea
and taking $ R\to \infty$.
In this limit the algebra \bref{SAdS} becomes
\bea
\{ Q, Q\}=-2iC\slP%{\cal P}_+
&~~,~~& 
\left[ Q ,P_{m}\right]=-\frac{i}{2}Z\Gamma_{m} 
\label{sptr}
%\begin{array}{lclcl}
%\left[ J_{mn},J_{lk} \right]=\eta_{[k|[m}J_{n]|l]}&,&
%\left[  Q ,J_{mn}\right]=-\frac{1}{2}Q\Gamma_{mn}&,&  
%\{ Q, Q\}=-2iC\slP%{\cal P}_+
%\\
%\left[ P_{m},J_{lk} \right]=\eta_{m[l}P_{k]}&,&
%\left[ Z ,J_{mn}\right]=-\frac{1}{2}Z\Gamma_{mn}&,&  
%\{Q,Z\}=0\\
%\left[ P_{m},P_{l} \right]=0&,&
%\left[ Q ,P_{m}\right]=-\frac{i}{2}Z\Gamma_{m}&,&  
%\{Z,Z\}=0\\
%&&
%%\left[ Z ,P_{m}\right]=0~.
%&&
%\end{array}\label{GS}
\eea 
with keeping correct Lorentz spins for the generators
\bea
\left[ J_{mn},J_{lk} \right]=\eta_{[k|[m}J_{n]|l]}&,&
\left[ P_{m},J_{lk} \right]=\eta_{m[l}P_{k]}\\
\left[  Q ,J_{mn}\right]=-\frac{1}{2}Q\Gamma_{mn}&,&  
\left[ Z ,J_{mn}\right]=-\frac{1}{2}Z\Gamma_{mn}\nn  .
\eea 
$Z_\alpha$ coincides with fermionic central charges 
in the supertranslation algebra \bref{sptr} introduced by Green \cite{Green}.

The algebra \bref{SAdS} 
is also written in a $SO(d-1,2)$ covariant notation as
\bea
\left[ J_{MN},J_{LK} \right]=\eta_{[K|[M}J_{N]|L]}~,~
\left[ {\cal Q} ,J_{MN}\right]=-\frac{1}{2}{\cal Q}\gamma_{MN}~,~
\{ {\cal Q}, {\cal Q}\}=C\gamma^{MN}J_{MN}~,~\label{LSAdS}
\eea
where indices $M$ run $M=\{m,d\}=\{0,1,\cdot\cdot\cdot,d-1,d\}$ and
\bea
J_{MN}&=&\{J_{mn}, J_{md}=P_m\}~,~\nn\\
{\cal Q}&=&\{{\cal Q}{\cal P}_+=Q, ~{\cal Q}{\cal P}_-=Z \}
\label{LSAdS1}\\
{\cal P}_\pm&=&\frac{1}{2}(1\pm i\gamma_d)~.\nn
\eea 
 
\vsej
%%%%%%%%%%%%%%%%%%%%%%%%%%%%%%%%%%%%%%%%%%%%%%%%%%%%%%%%%%%%%%%%%%%
\section{Maurer-Cartan 1-forms}
\indent

In this section we construct left invariant one forms of the coset group
G/H with G=(the nondegenerate super AdS group (\ref{SAdS})) and
H=(Lorentz group).
An element of the coset is parameterized as
\bea
g=g_Zg_Pg_Q=e^{\xi^\alpha Z_\alpha}e^{X^m P_m} e^{\theta^\alpha Q_\alpha}~~
\label{ggg}~~.
\eea
%since this parameterization provides simplest $\xi$ dependence in the flat %limit
%\cite{Sakag,AHKT}.
Maurer-Cartan 1-forms are defined as
\bea
\Omega=g^{-1}dg=L^A(X,\theta,\xi)T_A={\bf L}^mP_m+\frac{1}{2}{\bf L}_J^{mn}J_{mn}
+L^\alpha Q_\alpha+L_Z^\alpha Z_\alpha
\eea
and they satisfy the Maurer-Cartan equation $d\Omega=-\Omega^2$ 
whose components are
\bea
d{\bf L}^m&=&
	{\bf L}_J^{lm}{\bf L}_l
	+i\bar{L}\Gamma^m L-\frac{i}{R^2}\bar{L}_Z\Gamma^m L_Z\nn\\
d{\bf L}_J^{mn}&=&-\frac{1}{R^2}
	{\bf L}^{m}{\bf L}^n-{\bf L}_J^{lm}{\bf L}^n_{J~l}
	-\frac{2}{R^2}\bar{L}\Gamma^{mn} L_Z\label{MCeq}\\
d{L}^\alpha&=&-\frac{1}{4}{\bf L}_J^{mn}\Gamma_{mn}L
	+\frac{i}{2R^2}{\bf L}^{m}\Gamma_{m}L_Z
	\nn\\
d{L}_Z^\alpha&=&-\frac{1}{4}{\bf L}_J^{mn}\Gamma_{mn}L_Z
	-\frac{i}{2}{\bf L}^{m}\Gamma_{m}L
	\nn
\eea
with AdS radius or equivalently scaling parameter $R$. 

The 2-form potential and the 3-form field strength for a string are
given by
\bea
B=2i\bar{L}L_Z~~,~~
dB=\bar{L}\slbL L+\frac{1}{R^2}\bar{L}_Z\slbL L_Z~~\label{BandH}
\eea
using with \bref{MCeq}.

Next we will obtain expression of MC 1-forms of \bref{SAdS} where the scale parameter $R$ is not included. 
Although the parametrization \bref{ggg} leads following three kinds of
1-forms $L^A,{\cal L}^A,l'^A$
\bea
g_Z^{-1}d(g_Z)&=&l'=l'^A(\xi)T_A~~\nn\\
g_P^{-1}g_Z^{-1}d(g_Zg_P)&=&{\cal L}={\cal L}^A(X,\xi)T_A\label{dlDL}\\
g_Q^{-1}g_P^{-1}g_Z^{-1}d(g_Zg_Pg_Q)&=& L= L^A(X,\theta,\xi)T_A~~,\nn
\eea
the main structure of the total MC 1-forms is governed by $l^A$;
\bea
g_Q^{-1}dg_Q=l^A(\theta)T_A=d\theta^\alpha (l^A)_\alpha T_A~~.
\eea

The MC 1-forms are obtained as
\bea
({\bf L})^m&=&({\cal L}_P)^m
	+(l_P)^m_\alpha D{\theta}^\alpha
	+\frac{1}{8}(\bar{\theta}\Gamma^m\phi^{-4}\Gamma_{nl}\theta)
	(l_J)^{nl}_\alpha
	(\Upsilon_2 D'{\theta})^\alpha
\nn		\\
	&&+2i(\bar{\theta}\Gamma^m l_Q)_{\alpha}
	({\cal L}_Q)^\alpha
	 +\frac{i}{2}(\bar{\theta}\Gamma^m \Upsilon_2 \phi^{-4}l_Z
	 )_\alpha (\Upsilon_2 {\cal L}_Z)^\alpha
\nn\\
({\bf L}_J)^{mn}&=&({\cal L}_J)^{mn}
	+(l_J)^{mn}_\alpha D\theta^\alpha
	-\frac{1}{4}(\bar{\theta}\Gamma^{mn}\phi^{-4}\Upsilon_2 	
	\Gamma^l\theta)
	(l_P)_\alpha(D'\theta)^\alpha
\nn\\
	&&-2(\bar{\theta}\Gamma^{mn} l_Z)_\alpha
	({\cal L}_Q)^\alpha
	-(\bar{\theta}\Gamma^{mn}\Upsilon_1 
	l_Q)_\alpha (\phi^{-4}\Upsilon_2{\cal L}_Z)^\alpha
\nn\\
({ L})^\alpha&=&{\cal L}_Q^\alpha
	+(l_Q)_{~\beta}^{\alpha}D{\theta}^\beta
	+\frac{i}{8}(\Upsilon_2\phi^{-4}l_Z)^\alpha_{~\beta}
	(\Upsilon_2 D'{\theta})^{\beta} 
\label{LLL}\\
	&&-\frac{1}{4}
	(\Gamma_{mn}\theta)^\alpha
	(l_J)^{mn}_{\beta}{\cal L}_Q^\beta
	+\frac{i}{8}(\phi^{-4}\Upsilon_2\Gamma_m\theta)^\alpha
	(l_P)^m_\beta (\Upsilon_2 {\cal L}_{Z})^\beta
\nn	\\
({ L}_Z)^\alpha&=&{\cal L}_Z^\alpha
	+(l_Z)_{~\beta}^{\alpha}D{\theta}^\beta	
	+\frac{i}{4} (\Upsilon_1 l_Q)_{~\beta}^{\alpha}
	(\phi^{-4}\Upsilon_2 D'{\theta})^\beta-\frac{i}{2}
 	(\Gamma_m\theta)^\alpha
	(l_P)^m_\beta{\cal L}_Q^\beta	
\nn	 	\\
	&&+\frac{1}{8}
	(\Upsilon_1 \Gamma_{mn}\theta)^\alpha
	(l_J^{mn})_\beta (\phi^{-4}\Upsilon_2{\cal L}_Z	)^\beta\nn
 \eea
where the covariant derivatives on $\theta$ are
\bea
D\theta=d\theta+\frac{1}{4}({\cal L}_J)^{mn}\Gamma_{mn}\theta~~,~~
D'\theta=({\cal L}_P)^m\Gamma_m\theta~~.
\eea
In the above expressions $l^A, l'^A, \CL^A$ are defined respectively as
\bea
&&\left\{\begin{array}{lcl}
(l_P)^m&=&i\bar{\theta}\Gamma^m~(\cosh \phi- \cos\phi)\phi^{-2}~d\theta\\
(l_J)^{mn}&=&-\bar{\theta}\Gamma^{mn}~\Upsilon_1(\cosh \phi+ \cos\phi-2)\phi^{-4}~d\theta\\
(l_Q)^\alpha&=&\frac{1}{2}~
	\left((\sinh \phi +\sin\phi)\phi^{-1}\right)_{~\beta}^{\alpha}d{\theta}^\beta~\\
(l_Z)^\alpha&=&\frac{1}{2}~\left(\Upsilon_1
(\sinh \phi -\sin\phi)\phi^{-3}~\right)_{~\beta}^{\alpha}d{\theta}^\beta\\
\end{array}\right.\label{ltheta}\\
&&~~(\phi^4)^\alpha_{~\beta}=\frac{1}{2}(\Upsilon_2)^\alpha_{~\gamma} (\Upsilon_1)^\gamma_{~\beta}~,~
(\Upsilon_1)^\alpha_{~\beta}=(\Gamma^m\theta)^\alpha (\bar{\theta}\Gamma_m)_\beta~,~
(\Upsilon_2)^\alpha_{~\beta}=(\Gamma^{mn}\theta)^\alpha (\bar{\theta}\Gamma_{mn})_\beta~\nn
\eea
\bea
&&\left\{\begin{array}{lcl}
(l'_P)^m&=&-i\bar{\xi}\Gamma^m~(\cosh \phi'- \cos\phi')\phi'^{-2}~d\xi\\
(l'_J)^{mn}&=&-\bar{\xi}\Gamma^{mn}~\Upsilon'_1(\cosh \phi'+ \cos\phi'-2)\phi'^{-4}~d\xi\\
(l'_Q)^\alpha&=&\frac{1}{2}~
	\left(\Upsilon'_1
(\sinh \phi' -\sin\phi')\phi'^{-3}~	\right)_{~\beta}^{\alpha}d{\xi}^\beta\\
(l'_Z)^\alpha&=&\frac{1}{2}~\left(
(\sinh \phi' +\sin\phi')\phi'^{-1}\right)_{~\beta}^{\alpha}~d{\xi}^\beta
\end{array}\right.\label{ltheta'}\\
&&~~(\phi'^4)^\alpha_{~\beta}=\frac{1}{2}(\Upsilon'_2)^\alpha_{~\gamma} (\Upsilon'_1)^\gamma_{~\beta}~,~
(\Upsilon'_1)^\alpha_{~\beta}=(\Gamma^m\xi)^\alpha (\bar{\xi}\Gamma_m)_\beta~,~
(\Upsilon'_2)^\alpha_{~\beta}=(\Gamma^{mn}\xi)^\alpha (\bar{\xi}\Gamma_{mn})_\beta~~\nn
\eea
\bea
&&\left\{\begin{array}{lcl}
({\cal L}_P)^m&=&(
%\frac{
\sinh \sqrt{\Upsilon}%}{
/\sqrt{\Upsilon}%}
)^m_{~n}
	(dX^n-X_l(l'_J)^{ln})
	+(\cosh \sqrt{\Upsilon})^m_{~n}(l'_P)^n\\
({\cal L}_J)^{mn}&=&2X^m
	\left(
	%\frac{
	(1-\cosh \sqrt{\Upsilon})%}{
	/\Upsilon)%}
	(dX-Xl'_J )\right)^n
	+(l'_J)^{mn}\\
	&&~-2X^m\left((%\frac{
	\sinh\sqrt{\Upsilon}%}{
	/\sqrt{\Upsilon}%}
	)l'_P\right)^n\\
({\cal L}_Q)^\alpha&=& \cosh (\slX/2)_{~\beta}^{\alpha}(l'_Q)^\beta
	+i\sinh(\slX/2)_{~\beta}^{\alpha}(l'_Z)^\beta\\
	({\cal L}_Z)^\alpha&=& \cosh (\slX/2)^\alpha_{~\beta}(l'_Z)^\beta
	-i\sinh(\slX/2)^\alpha_{~\beta}(l'_Q)^\beta
\end{array}\right.\label{Ltheta}\\
&&~~\Upsilon^{mn}=\eta^{mn}X^2-X^mX^n~~~~~~~~~~~~~~~~~~~~~~~~~~~~~~~~~~~~~~~~~~~~~~~~~~~~~~~~~~~~~~~~~~~~~~~~~~\nn
\eea
where $l^A$ and $l'^A$ are written as four-module functions
\footnote{
\bea
\sum_{n=0}\frac{\phi^{4n}}{(4n)!}=\frac{1}{2}(\cosh \phi +\cos \phi)
&,&
\sum_{n=0}\frac{\phi^{4n+1}}{(4n+1)!}=\frac{1}{2}(\sinh \phi +\sin \phi)
\nn\\
\sum_{n=0}\frac{\phi^{4n+2}}{(4n+2)!}=\frac{1}{2}(\cosh \phi -\cos \phi)
&,&
\sum_{n=0}\frac{\phi^{4n+3}}{(4n+3)!}=\frac{1}{2}(\sinh \phi -\sin \phi)~~,
\label{cosin}
\eea
}
rather than usual two-module functions
and ${\cal L}_P$ and ${\cal L}_J$ are usual vielbein and connection
respectuvely.

%%%%%%%%%%%%%%%%%%%%%%%%%%%%%%%%%%%%%%%%%%%%%%%%%%%%%%%%
\section{Superstring in the nondegenerate super-AdS space}
\indent

The nondegenerate algebra $T_A=(Q_\alpha, P_m, Z_\alpha)$
given by \bref{SAdS} 
enables to introduce 
nondegenerate group metric \cite{Green,Siegel}
\bea
tr(T_AT_B)=\left(\begin{array}{ccc}
		0&0&C_{\alpha\beta}\\
		0&\frac{1}{2}\eta_{mn}&0\\
		-C_{\alpha\beta}&0&
		\end{array}		
		\right)~~.
\label{metric}
\eea
A superstring action 
can be given in the following form using with
 MC 1-forms 
$g^{-1}dg=L=d\sigma^\mu L_\mu=d\sigma^\mu L_\mu^{~A} T_A$ 
obtained as
\bref{LLL}
\bea
S&=&S_0+S_{WZ}\nn\\
S_0&=&-T~{\rm tr}\int d^2\sigma [ h^{\mu\nu}L_\mu L_\nu]
=-\frac{T}{2}~\int d^2\sigma h^{\mu\nu}
{\bf L}_\mu^m {\bf L}_{\nu m}\label{action0}\\
S_{WZ}&=&~T~{\rm tr}\int d^2\sigma [\epsilon^{\mu\nu}L_\mu L_\nu]
=~2T~\int d^2\sigma \epsilon^{\mu\nu}\bar{L}_\mu  L_{Z \nu }\label{actionwz}
\eea
with $h^{\mu\nu}=\sqrt{-g}g^{\mu\nu}$ and $\epsilon^{01}=1$.
Following to \bref{scaleAdS} these variables are scaled as
$X\to (1/R)X$,
 $\theta \to (1/\sqrt{R})\theta$
 and $\xi\to (1/\sqrt{R}^3)\xi$
and also
${\bf L}\to (1/R){\bf L}$,
 $L \to (1/\sqrt{R})L$
 and $L_Z\to (1/\sqrt{R}^3)L_Z$.
In the limit $R\to \infty$ they reduce to the MC 1-forms in the flat space:
\bea
&&\left\{\begin{array}{lcl}
({\bf L})_\mu^m&=&\partial_\mu X-i\bar{\theta}\Gamma \partial_\mu\theta\nn\\
&&+\frac{1}{R^2}\left(
	\frac{1}{3!}\Upsilon^{mn} \partial_\mu X_n
	+i\bar{\xi}\Gamma^m\partial_\mu\xi
	+\bar{\theta}\Gamma^m\slX\partial_\mu\xi
	+\frac{i}{4}\bar{\theta}\Gamma^{mnl}\theta X_n\partial_\mu X_l
\right)\\
&&+o(\frac{1}{R^4})\nn\\
({ L})_\mu^\alpha&=&\partial_\mu\theta+\frac{1}{R^2}\left(
	\frac{i}{2}\slX\partial_\mu \xi
	-\frac{1}{4}X^m \partial_\mu X^n\Gamma_{mn}\theta
	+\frac{1}{4}\Upsilon_2\partial_\mu\xi
\right)+o(\frac{1}{R^4})	\label{LLLexp}\\
({ L}_Z)_\mu^\alpha&=&
	\partial_\mu \xi^\alpha+\frac{i}{2}(\Gamma_m\theta)^\alpha
	(\partial_\mu X^m-\frac{i}{3}\bar{\theta}\Gamma^m \partial_\mu\theta)\nn\\
	&&+\frac{1}{R^2}\left(
	-\frac{1}{2}(\bar{\xi}\Gamma \partial_\mu \xi)\cdot\Gamma\theta
	 +\frac{1}{2}(\frac{\slX}{2})^2\partial_\mu \xi
	 +\frac{i}{4}\Upsilon_1\slX\partial_\mu\xi
	+\frac{i}{2\cdot 3!}(\Upsilon\cdot \partial_\mu X)\cdot\Gamma\theta
	\right)\nn\\
	&&+o(\frac{1}{R^4})\nn
\end{array}\right.\\&&
 	\label{LLLexpR}~~
\eea
where $\theta^3$ and $\xi^3$ vanish in 3-dimensional AdS space.

The supertransformation rules which leave \bref{LLLexpR} invariant
up to Lorentz rotation
are determined independently of the action.
They are calculated by performing the infinitesimal supertransformation on the coset element
\bea
&&g~~\to~~e^{\epsilon Q}g=g'h~~,~~h\in H\nn\\
&&~~\left\{\begin{array}{lcl}
g&=&e^{\xi Z}e^{X\cdot P}e^{\theta Q}\\
g'&=&e^{(\xi+\delta \xi) Z}e^{(X+\delta X)\cdot P}e^{(\theta+\delta \theta) Q}
\end{array}\right.~~~~~.
\eea
The obtained N=1 AdS$_3$ supertransformation rules for large $R$ are
\bea
\left\{\begin{array}{lcl}
\delta_\epsilon X^m&=&-i\bar{\theta}\Gamma^m\epsilon +\frac{1}{R^2}\left(
-\frac{X^2}{8}i\bar{\theta} \Gamma^m\epsilon+\frac{X^2}{6}\Upsilon^m_n
i\bar{\theta}\Gamma^n \epsilon+\frac{1}{2}\bar{\xi}\Gamma^m\slX\epsilon
\right)+o(\frac{1}{R^4})\\
\delta_\epsilon \theta&=&\epsilon+\frac{1}{R^2}\left(
-\frac{X^2}{8}\epsilon+\frac{1}{2}(\bar{\xi}\Gamma^{mn}\epsilon \Gamma_{mn}\theta)
\right)+o(\frac{1}{R^4})\label{SUSYrule}\\
\delta_\epsilon \xi&=&-\frac{i}{2}\slX \epsilon
-\frac{1}{6}\Gamma\theta\cdot \bar{\theta}\Gamma\epsilon
+\frac{1}{R^2}\left(
-\frac{X^2}{48}i\slX\epsilon
\right)+o(\frac{1}{R^4})~~~~
\end{array}\right.~~.
\eea
The supercharges are obtained independently from 
the form of the Wess-Zumino action in this approach:
\bea
Q\epsilon=\int \left(\zeta \delta_\epsilon \theta+
p\delta_\epsilon X+\pi_\xi\delta_\epsilon \xi
\right)
\eea 
where $(X,\theta,\xi)$ are canonical variables and $(p,\zeta,\pi_\xi)$ are 
their conjugates.
By construction they satisfy the following superalgebra \bref{SAdS}
\bea
\{Q,Q\}=-2iC\slP~~.
\eea

In the nondegenerate approach, the Wess-Zumino action does not affect the supercharges and the superalgebra,
but does affect the fermionic constraints and their anti-commutator.
The canonical conjugates scale are defined as
%as 
%$p_m \to R p_m$ and $\zeta_\alpha \to \sqrt{R}\zeta_\alpha$;
\bea
\zeta_\alpha&\equiv& \frac{\delta^r S}{\delta \dot{\theta}^\alpha}
=\frac{\delta S_0}{\delta {\bf L}_0}\cdot
	\frac{\partial^r {\bf L}_0}{\partial \dot{\theta}^\alpha}
+\frac{\delta^r S_{WZ}}{\delta \dot{\theta}^\alpha}\label{condef}\\
p_m&\equiv&\frac{\delta S}{\delta \dot{X}^m}=\frac{\delta S_0}{\delta {\bf L}_0}\cdot
	\frac{\partial {\bf L}_0}{\partial \dot{X}^m}
+\frac{\delta S_{WZ}}{\delta \dot{X}^m}
%&=& \tilde{p}\cdot\sum \frac{1}{R^{2N}}\frac{\partial^r {\bf L}_0^{(N)}
%}{\partial \dot{\theta}}
%+\sum \frac{1}{R^{2N}}\frac{\delta^r {S}_{WZ}^{(N)}
%}{\delta \dot{\theta}}
\eea
where $\delta^r$ denotes the right derivative. 
%It is convenient to introduce $\tilde{p}$ which 
%is SUSY invariant combination 
%of $p$ 
%\bea
%\tilde{p}=p+\frac{\delta {S}_{WZ}}{\delta \dot{X}}
%%=p- \sum \frac{1}{R^{2N}}\frac{\delta {S}_{WZ}^{(N)}}{\delta \dot{X}}~~.
%\eea
%Combining the above expressions gives
From the definition \bref{condef}
 fermionic constraints are written as
\bea
F&=&\sum_N \frac{1}{R^{2N}}F^{(N)}%=
%\sum_N\frac{1}{R^{2N}}(f_0^{(N)}+f_{WZ}^{(N)})
~=~0%\\
%f_0^{(0)}&=&\zeta-p\cdot \frac{\partial^r {\bf L}_0^{(0)}}{\partial %\dot{\theta}}
%\nn\\
%f_0^{(N)}&=&-p\cdot\frac{\partial^r {\bf L}_0^{(N)}}{\partial \dot{\theta}}
%\nn~~~~(N\neq 0)\nn\\
%f_{WZ}^{(N)}&=&\sum_{L=0}^N
%\frac{\delta S_{WZ}^{(N-L)}}{\delta \dot{X}}
%\frac{\partial^r {\bf L}_0^{(L)}}{\partial \dot{\theta}}
%-\frac{\delta^r S_{WZ}^{(N)}}{\delta \dot{\theta}}~~,\label{F}
%\nn
\eea
which satisfy the anti-commutator of the fermionic constraints
\bea
\{F_\alpha(\sigma),F_\beta(\sigma')\}
%=2i(C\tilde{\slp})_{\alpha\beta}+\cdot\cdot\cdot
=2i(C{\slp})_{\alpha\beta}+\cdot\cdot\cdot
\eea
and the terms ``$\cdot\cdot\cdot$" will be calculated in the following sections.
The fermionic local constraints
are SUSY invariant.

%%%%%%%%%%%%%%%%%%%%%%%%%%%%%%%%%%%%%%%%%%%%%%%%%%%%%%%%%%%%%%%

\par
\vsej
\subsection{Flat case ($1/R~\to~ 0$)}
\indent

In the flat case the Wess-Zumino Lagrangian for nondegenerate superspace
approach is given by
\bea
{\cal L}_{WZ}&=&{\cal L}_{WZ,GS}+\triangle {\cal L}\label{wessz}\\
{\cal L}_{WZ,GS}&=&T\epsilon^{\mu\nu}(
i\partial_\mu\bar{\theta}\Gamma\theta\cdot\partial_\nu X
-\frac{1}{3}\partial_\mu\bar{\theta}\Gamma\theta\cdot
\partial_\nu\bar{\theta}\Gamma\theta)
\nn\\
\triangle {\cal L}_{WZ}&=&2T\epsilon^{\mu\nu}\partial_\mu 
\bar{\theta}\partial_\nu \xi\label{WZs}\quad,\nn
\eea 
where ${\cal L}_{WZ,GS}$ is the
 Wess-Zumino Lagrangian for the Green-Schwarz superstring
and $\triangle{\cal L}_{WZ}$ is rewritten in total derivative form.
Since $\xi$-dependence is only in the surface term,
 obviously $\xi$ is not dynamical. 
However $\xi$ is transformed in such a away that the Wess-Zumino action is 
 invariant under the SUSY transformations. 
The supercharges and the superalgebra are given as
\bea
&&Q\epsilon =\label{scnd}\int d\sigma \left(
\zeta-i\bar{\theta}\slp-\frac{i}{2}\pi_\xi \slX-\frac{1}{6}\pi_\xi\Gamma\theta\cdot\bar{\theta}\Gamma
\right)\epsilon\\
&&P_m=\int p_m~~,~~
Z_\alpha=\int \pi_{\xi,\alpha} \label{SUSYnd}\nn\\
&&
\{Q,Q\}=-2iC\slP~~,~~[Q,P_m]=-\frac{i}{2}Z\Gamma_m~~\label{ndsc}
~~~~.
\eea
In flat case $Z$ is a center, $\{Z,Z\}=\{Z,Q\}=[Z,P]=0$.

On the other hand, the usual GS supercharges and their SUSY algebra 
are given by
\bea
&&Q_{GS}\epsilon =\int \left(
p\delta_\epsilon X+\zeta \delta_\epsilon \theta -U_{WZ}^0\right)~~~~~~~,~~
\delta_\epsilon {\cal L}_{WZ,GS}=\partial_\mu U^\mu_{WZ}\label{scgs}\\
&&~~~~=\int\left(
\zeta-i\bar{\theta}\slp-T(
i\bar{\theta} \slX'+\frac{1}{3}\bar{\theta}'\Gamma\theta\cdot\bar{\theta}\Gamma)
\right)\epsilon\nn\\
&&\nn ~\Sigma_m=T\int X'_m~\\
&&\{Q_{GS},Q_{GS}\}=-2iC(\slP+\slSigma)~~,~~\label{GSsc}
[Q_{GS},P_m]=0~
~~.
\eea
The presence of ``$\xi$" leads to the difference of superalgebras
\bref{ndsc} and \bref{GSsc}.
Especially the topological term appears 
in the usual GS superalgebra but not in the nondegenerate superalgebra.

In the nondegenerate approach, the existence of the Wess-Zumino action does not affect the supercharges and the superalgebra,
but does affect the fermionic constraints and their anti-commutator.
The fermionic constraint set and its algebra are
\bea
&&F=F^{(0)}=(\zeta+i\bar{\theta}\slp)+
T(\bar{\theta}\Gamma\theta'\cdot\bar{\theta}\Gamma-2\bar{\xi}'
+i\bar{\theta}\slX')=0\label{fercon}\\
&&F_Z=\pi_\xi+2T\bar{\theta}'=0\label{ferconZ}\\
&&\{F_\alpha(\sigma),F_\beta(\sigma')\}=
2iC\Gamma\cdot(\tilde{p}+T{\bf L}_1^{(0)})\delta(\sigma-\sigma')
\label{susyff}
\\
&&\{F_{Z,\alpha}(\sigma),F_{Z,\beta}(\sigma')\}=0\label{susyffz}
~=~\{F_\alpha(\sigma),F_{Z,\beta}(\sigma')\}
\eea
where $\tilde{p}$ is SUSY invariant combination given by 
\bea
\tilde{p}=p+iT\bar{\theta}'\Gamma\theta~~
\eea
and ${\bf L}_1^{(0)}$ is given by \bref{LLLexpR}.
The anti-commutator of fermionic constraints \bref{susyff}
is the same as one of the Green-Schwarz.

The relation of two supercharges becomes clear by imposing
 constraints in \bref{fercon} 
 $F_Z=0~\to~\pi_\xi=-2T\bar{\theta}'$,
\bea
Q\epsilon|_{F_Z=0}=Q_{GS}\epsilon-
2T\int \partial (\bar{\theta}\delta_\epsilon \xi)~~.\label{GSnddiff}
\eea
For a case of an open string
the surface term does not vanish ,
$ \bar{\theta}\slX\epsilon|_{\sigma=0}^{\sigma=\pi}\neq 0$,
and it causes breaking of translational invariance giving the $Z$ charge.
For cases such as a closed string and a string with the 
periodic boundary condition 
the surface term and the fermionic charge $Z$ vanish.

Topological charge is an important issue.
 Although the SUSY algebra does not contain a topological term,
the action and the local fermionic constraints contain
the topological term information. 
If we impose $F=0$ of \bref{fercon}
in supercharges
\bea
Q\epsilon|_{F=0}=Q\epsilon-\int F\epsilon
=\int \left(
2p\delta_\epsilon X +\pi_\xi \delta_\epsilon \xi
-F_{WZ}\epsilon \right)~~
\eea
where $F_{WZ}$ is $T$-dependent part (i.e. the Wess-Zumino action dependent part) of $F$,
they produce the topological term in their 
bracket 
\bea
\{Q|_{F=0},Q|_{F=0}\}&=&2iC\slSigma \label{Qstar}~~.
\eea
This is a result of the fact that $\{Q,F\}=0$
and the anti-commutator of the fermionic constraints
carry the topological information, $\{\int F,\int F\}
=2iC(\slP+\slSigma)$.

%\bea
%\{Q|_{F=0},Q|_{F=0}\}&=&\{Q,Q\}+\{\int F,\int F\}-\{Q,\int F\}-\{\int %F,Q\}\nn\\
%&=&\left(-2i C\slP\right)+2iC(\slP+\slSigma)~~.
%\eea

It is stressed that $Q|_{F_Z=0}$ and $Q|_{F=0}$ are not conserved 
Noether charges of the system.
Second class parts of $F_Z=0$ and $F=0$ play a role of leading to
 relations \bref{GSnddiff}
 and \bref{Qstar}. 

%%%%%%%%%%%%%%%%%%%%%%%%%%%%%%%%%%%%%%%%%%%%
\medskip
\subsection{AdS case (finite $R$)}
\indent

In this section we calculate the SUSY algebra and the anti-commutator of the fermionic constraints in next to leading order of $1/R$. 
Then we generalize this result for finite $R$ expression,
comparing with the usual GS superstring case.
It is straightforward to confirm the SUSY algebra \bref{SUSYnd}
in the next to leading order by using \bref{SUSYrule}.
Next we calculate the anti-commutator of the fermionic constraints.
The fermionic constraints are obtained as
\bea
F&=&F^{(0)}+\frac{1}{R^2}F^{(1)}=0~~\label{ferconRR}\\
F^{(1)}&=&
-\frac{i}{3!}p_m\Upsilon^{mn}
\bar{\theta}\Gamma_n
	+T\left\{\right.
	i\bar{\theta}\Gamma_m(
	\frac{i}{3}\Upsilon^{mn}\bar{\theta}\Gamma_n\theta'
	+\frac{1}{6}\Upsilon^{mn}X'_n\nn\\
&&	+\frac{1}{2}X^m\bar{\xi}'\theta
	+i\bar{\xi}\Gamma^m\xi'
	+\frac{2}{3}\bar{\theta}\Gamma^m\slX \xi')
-\frac{1}{4}\bar{\xi}'\slX^2
\left.\right\}
~~~~\nn
\eea
where $F^{(0)}$ is given in \bref{fercon}.
The anti-commutator of the fermionic constraints in next to leading order is
\bea
\{F_\alpha(\sigma),F_\beta(\sigma')\}&=&2i(C\Gamma)_{\alpha\beta}\cdot
(\tilde{p}+T{\bf L}_1)\delta(\sigma-\sigma')
\label{localsusy}
\eea
with $\tilde{p}$ given by
\bea
\tilde{p}_m&=&p_m+Ti\bar{\theta}'\Gamma_m \theta
\nn\\
&&+\frac{1}{R^2}
\left[\right.
-\frac{1}{3!}\Upsilon_{mn}p^n
-\frac{i}{8}\bar{\theta}\Gamma_{mnl}\theta X^np^l\nn\\
&&+T\left(
-\frac{i}{8}\bar{\theta}\Gamma_{mnl}\theta X^nX'^l
+\frac{i}{3}\Upsilon_{mn}\bar{\theta}\Gamma^n\theta'
+\frac{1}{4}\bar{\theta}\slX\Gamma_m\xi'
\right)\left.\right]\nn\\
\eea
and ${\bf L}_1={\bf L}_1^{(0)}+\frac{1}{R^2}{\bf L}_1^{(1)}$ given in \bref{LLLexpR}.
$\tilde{p}$ and ${\bf L}_1$ are SUSY invariant combinations.
From the SUSY invariance the form of the anti-commutator of the fermionic constraints,
\bref{localsusy} should be hold 
in all order of the $1/R$ expansion.
The usual GS superstring case 
the same form of the
anti-commutator of the fermionic constraints \bref{localsusy} is expected,
although the concrete expression of the fermionic constraints $F=0$ is
completely different.
This anti-commutator of the fermionic constraints \bref{localsusy} guarantees the $\kappa$-symmetry of the system
for both the GS superstring and the nondegenerate superstring:
The rank of the matrix in the
right hand side of \bref{localsusy} is half of its dimension.
 Since 
$(\tilde{p}+T{\bf L}_1)$ is light-like vector,
$(\tilde{p}+T{\bf L}_1)^2\approx 0$ by using with the diffeomorphism
constraints, $(\tilde{p}+T{\bf L}_1)_m\Gamma^m$ is a projection 
operator.
Projected constraints $F\Gamma\cdot (\tilde{p}+T{\bf L}_1) =0$ are
first class constraints generating $\kappa$-symmetry.

Analogous to the flat case \bref{Qstar}, the global SUSY algebra will produce 
a brane charge if the fermionic constraints
$F=0$ of \bref{ferconRR} is used in $Q$,
\bea
\{(Q|_{F=0})_\alpha,(Q|_{F=0})_\beta\}&=&-2iT(C\Gamma)_{\alpha\beta}\cdot\int d\sigma{\bf L}_1
\nn\\&=&-2iT(C\Gamma)_{\alpha\beta}\cdot\int d\sigma ~\sinh \sqrt{\Upsilon}
\cdot  {\Upsilon}^{-1/2}\cdot X'\quad .\label{tpltrm}
\eea
The BPS states are eigenstates of this brane charge.

For finite $R$, $Z$ is not center any more.
 $Q$ and $Z$ satisfy 
almost same algebra except their opposite sign of the right hand side in \bref{SAdS}.
The opposite sign of $\{Q,Q\}$ and $\{Z,Z\}$ comes from the dimensional reduction from the larger algebra
\bref{LSAdS} and \bref{LSAdS1}.
Since the square of the supercharges leads to positivity of the energy,
the new supercharge $Z$ looks the wrong sign for unitarity.
In order to make the algebras in \bref{SAdS} to be consistent,
$Z$ must be anti-hermite:
The algebras are represented on physical states
as $[Q,Q]_+=2E=-[Z,Z]_+$. 
If $Q$ is hermite then
the algebra of two $Q$'s is consistent with positivity of the energy
$\langle\psi|QQ|\psi\rangle=\sum_n|\langle\psi|Q|n\rangle|^2
=E_\psi\langle\psi|\psi\rangle
\geq 0$. 
On the other hand, positivity requires
$0\leq  E_\psi\langle\psi|\psi\rangle
=\sum_n|\langle\psi|Z|n\rangle|^2$,
and the algebra of $Z$ leads to
$
-\langle\psi|ZZ|\psi\rangle= 
-\sum_n\langle\psi|Z|n\rangle \langle n|Z|\psi\rangle 
=-\sum_n\langle\psi|Z|n\rangle (\langle\psi|Z^\dagger|n\rangle )^*
$, therefore $Z^\dagger=-Z$.
The anti-hermiticity of $Z$ is consistent with 
\bref{SAdS} because
there can be pure real Majorana representation in 3-dimension.
It requires replacing
 $\xi\to i\xi$ and $\bar{\xi}=\xi^{T}C\to i\bar{\xi}$.

Now for a case with $R=1$, 
$\theta$ and $\xi$ are treated as same and generate N=2 supersymmetry.
As expected the $\kappa$ symmetry for $\xi$ is guaranteed by   
 the fermionic constraints for $\xi$, $F_Z=0$;
\bea
\{F_{Z,\alpha}(\sigma),F_{Z,\beta}(\sigma')\}=
-2i(C\Gamma)_{\alpha\beta}\cdot(\tilde{p}-T{\bf L}_1)\delta(\sigma-\sigma')
\quad .\label{localsuz}
\eea
Therefore dynamical modes of the nondegenerate AdS superstring model
is twice of the usual Green-Schwarz's one.
For general finite $R$ case is obtained analogously by rescaling. 

It is curious that 
dynamics of $\theta$ and $\xi$ look the same 
in a case for AdS with $R=1$,
but they are not the same in the flat limit.
In order to see the difference of their dynamics,
let us compare the equation of motion in the next to leading order:
\bea
\frac{\delta S}{\delta \theta}&=&-
[2\left((1+\frac{1}{R^2}\frac{1}{3!}\Upsilon)\partial_\mu X \right)^+
(g^{\mu\nu}-\epsilon^{\mu\nu})i\partial_\nu\bar{\theta}\Gamma^-\nn\\
&&~~+\partial_\nu\{\left((1+\frac{1}{R^2}\frac{1}{3!}\Upsilon)\partial_\mu X \right)^+
(g^{\mu\nu}-\epsilon^{\mu\nu})\}i\bar{\theta}\Gamma^-]\nn\\
&&+\frac{1}{R^2}[
2(g^{\mu\nu}-\epsilon^{\mu\nu})(\bar{\xi}\Gamma^+\partial_\nu \xi)
\partial_\mu \bar{\theta}\Gamma^-
+\partial_\mu\{
(g^{\mu\nu}-\epsilon^{\mu\nu})\bar{\xi}\Gamma^+\partial_\nu \xi \}
\bar{\theta}\Gamma^-\nn\\&&
~~-\frac{1}{2}(g^{\mu\nu}-\epsilon^{\mu\nu})\partial_\mu X^2 
\partial_\nu \bar{\xi}]~=~0\\
\frac{\delta S}{\delta \xi}&=&-\frac{1}{R^2}[
2(g^{\mu\nu}+\epsilon^{\mu\nu})(\partial_\mu X^--i\bar{\theta}\Gamma^-\partial_\mu
\theta)i\partial_\nu\bar{\xi}\Gamma^+\nn\\
&&~~+\partial_\nu\{
(g^{\mu\nu}+\epsilon^{\mu\nu})(\partial_\mu X^--i\bar{\theta}\Gamma^-\partial_\mu
\theta)\}i\bar{\xi}\Gamma^+\nn\\
&&~~+\frac{1}{2}\partial_\nu\{
(g^{\mu\nu}+\epsilon^{\mu\nu})\partial_\mu X^2\bar{\theta}
\}]~=~0~~~,
\eea  
where these fermionic variables are gauge fixed as
$\Gamma^+\theta=\Gamma^-\xi=0$ for simple comutation.
In the flat limit $\theta$ satisfies 
$(\partial_0-\partial_1)\theta=0$ and the equation for $\xi$ does not exist
as expected.
In the next to leading order,
the AdS effect appears in the equation for $\theta$,
and the equation of motion for $\xi$ appears
then $\xi$ becomes dynamical.
%In the limit $R=1$ dynamics of $\theta$ and $\xi$
%become the same by construction.    

%%%%%%%%%%%%%%%%%%%%%%%%%%%%%%%%%%%%%%%%%%%%%

\section{Summary and discussions}\par
\indent

We have extended the nondegenerate supertranslation algebra to the one in
AdS space.
In flat space the fermionic central extension is the way to make 
a superspace to be nondegenerate,
but it is not so in AdS space.
Introducing partner supercharges $Z$ is 
essential to make the superspace to be nondegenerate
irrespective of its central property
as we have shown in this paper.
The algebra \bref{SAdS} is almost unique
which contain minimal set of 
the super AdS  generators plus additional supercharges
and reproduces the fermionic central extended form in flat limit.
Further extension, such as extended SUSY and other target spaces
like AdS$\times$S,
of the nondegenerate super-AdS algebra 
are non-trivial problems.
The difficulty is
that more fermionic charges require more 
bosonic charges which can not be identified with
spacetime symmetry generators nor
R-symmetry generators.
In other words
nondegenerate SUSY partners may be just elements in 
the large multiplet accompanied with 
brane charges \cite{Sez,Sakag,AHKT,Sakag2,azc1}.
 Interpretation of $Z$ as the fermionic brane charge 
 recently examined \cite{HS} may be essential.

We examined the new superstring action based on this 
nondegenerate super-AdS algebra:
The difference of the nondegenerate approach from the usual GS superstring is clarified.
The supercharge of the nondegenerate approach is
related to the ones of GS superstring as \bref{GSnddiff},
and this surface term leads to 
difference in the global charge algebras
\bref{ndsc} and \bref{GSsc}.
While the fermionic charge $Z$ is center in the flat space, 
$Z$ generates another SUSY in the AdS space as discussed in section 4.2.
The equation \bref{tpltrm}
gives concrete expression of the brane charge in AdS space
which should appear in the superalgebra of the usual
GS superstring in AdS.
 The origin of the nontrivial relation between our action and the 
Green-Schwarz action is the ambiguity of the new fermion's scaling weight
in the AdS space where the scale parameter exists. 
For example; \\
(i) Conventional choice 
\begin{center}
\begin{tabular}{c|c|l}
Charges in AdS space &	scaling weight	&Charges in flat space	\\
\hline
$	P_m	$&   1	&~~~$P_m$ ...survived\\
		$Q_\alpha$&   1/2	&~~~$Q_\alpha$ ...survived\\         
	$	Z_\alpha$  &1/2&~~~$Z_\alpha$ ...survived 
\end{tabular}
\end{center}
Conventional choice of the scaling for spinors gives
the N=2 GS variables, but 
the bilinear form  Wess-Zumino action vanishes 
in the flat limit.
\\
(ii) Our choice \bref{scaleAdS}
\begin{center}
\begin{tabular}{c|c|l}
Charges in AdS space &	scaling	weight&	Charges in flat space	\\
\hline
$		P_m$&	   1&~~~$P_m$ ...survived\\
$		Q_\alpha$&   1/2&~~~$Q_\alpha$ ...survived\\
$		Z_\alpha $&  3/2&~~~$Z_\alpha$ ...center(auxiliary)
\end{tabular}
\end{center}
This choice gives the N=1 GS variables plus an auxiliary spinor
in the flat limit, 
and a simple bilinear Wess-Zimino action can be constructed 
even in the flat limit.

The nondegenerate superstring should be useful for quantum theory,
since fermionic states have their nondegenerate norm keeping 
canonical pairs $\theta$ and $\xi$.
In the AdS space the new fermion $\xi$ is dynamical as same as $\theta$
and both modes contribute to the Hamiltonian.
In the flat limit, although $\xi$ dependence is just in a surface term
in the classical action,
there exist the quantum states of $\xi$ 
making nondegenerate metric for original fermion $\theta$
-only Hamiltonian hides $\xi$ dependence-.
The ``nondegenerate" approach could give the  
formulation of the random superstring \cite{Siegel} 
which is worth to translate
into the continuum quantization theory.
The ``nondegenerate" approach manifests symmetrical structure of the system in the continuum action too, so it will be also useful for the continuum quantizaton. 
Further studies are required to clarify these issues.

\vskip 6mm
{\bf Acknowledgments}\par
This research is partially supported by the Grant-in-Aid for Scientific 
Research, \\\noindent
No.12640258 Ministry of Education Japan.
%\appendix
%\eject
%%%%%%%%%%%%%%%%%%%%%%%%%%%%%%%%%%%%%%%%%%%%%%%%%%%%%%%%%%%%%%%%%%

\end{document}